\documentclass{article}
\usepackage{ijcai19}

\usepackage{enumitem}
\usepackage{times}
\usepackage{soul}
\usepackage{url}
\usepackage[hidelinks]{hyperref}
\usepackage[utf8]{inputenc}
\usepackage[small]{caption}
\usepackage{graphicx}
\usepackage{amsmath}
\usepackage{mathtools}
\usepackage{booktabs}
\usepackage{algorithm}
\usepackage{algorithmic}
\usepackage{multirow}
\usepackage{amssymb}

\makeatletter
\def\hlinew#1{%
	\noalign{\ifnum0=`}\fi\hrule \@height #1 \futurelet
	\reserved@a\@xhline}

\title{Explainable Fashion Recommendation: \\A Semantic Attribute Region Guided Approach}

\author{
Min Hou$^{1,2}$\and
Le Wu$^3$\and
Enhong Chen$^{1,2}$\footnote{Corresponding Author}\and
Zhi Li$^{1,2}$\and
Vincent W. Zheng$^4$\And
Qi Liu$^{1,2}$\\
\affiliations
$^1$Anhui Province Key Lab. of Big Data Analysis and Application, University of S\&T of China\\
$^2$School of Data Science, University of S\&T of China\\
$^3$Hefei University of Technology\\
$^4$WeBank\\
\emails
\{minho, zhili03\}@mail.ustc.edu.cn,
lewu@hfut.edu.cn,
\{qiliuql, cheneh\}@ustc.edu.cn,
vincentz@webank.com
}

\begin{document}

\maketitle

\begin{abstract}

 In fashion recommender systems, each product usually consists of multiple semantic attributes (e.g., sleeves, collar, etc). When making cloth decisions, people usually show preferences for different semantic attributes~(e.g., the clothes with v-neck collar). Nevertheless, most previous fashion recommendation models comprehend the clothing images with a global content representation and lack detailed understanding of users' semantic preferences, which usually leads to inferior recommendation performance. To bridge this gap, we propose a novel \textit{S}emantic \textit{A}ttribute \textit{E}xplainable \textit{R}ecommender \textit{S}ystem (SAERS). Specifically, we first introduce a fine-grained interpretable semantic space. We then develop a Semantic Extraction Network (SEN) and Fine-grained Preferences Attention (FPA) module to project users and items into this space, respectively.  With SAERS, we are capable of not only providing cloth recommendations for users, but also explaining the reason why we recommend the cloth through intuitive visual attribute semantic highlights in a personalized manner. Extensive experiments conducted on real-world datasets clearly demonstrate the effectiveness of our approach compared with the state-of-the-art methods.
  
\end{abstract}

\section{Introduction}

The ubiquity of online fashion shopping has led to information explosion in the fashion industry. That imposes an increasing challenge for users who have to choose from a large number of available fashion products for satisfying personalized demands. Moreover, to promote profits growth, the fashion retailers have to understand the preferences of different customers and provide more intelligent recommendation services. However, unlike generic objects, clothes usually present significant variations of visual appearance, which has a vital impact on consumer decision.
\begin{figure}
    \centering
	\includegraphics[width=0.45\textwidth,height=100pt]{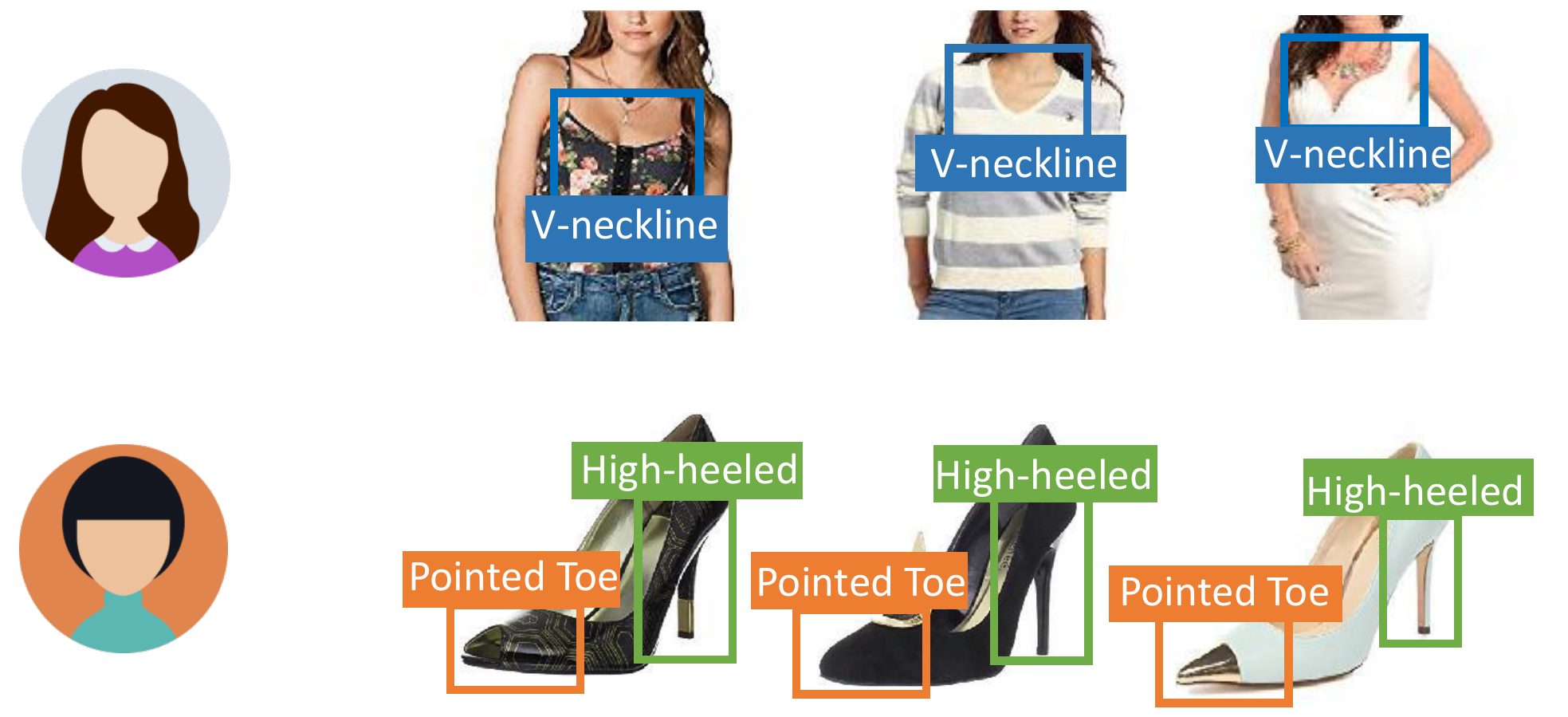}
	\caption{An example of user preferences for semantic attributes.}
    \centering
    \bigskip
    \label{fig:intro}
\end{figure}

Along this line, some studies have enhanced the performance of fashion recommendation via various visual features. For example, ImageNet\footnote{\url{http://image-net.org/}} pre-trained convolutional neural networks (CNNs) was adopted to extract visual representation~\cite{mcauley2015image,he2016vbpr,he2016sherlock}. As these visual features were trained for the image classification task, the pre-trained CNNs could only capture category information. Yu \textit{et al.}~\shortcite{yu2018aesthetic} and Liu \textit{et al.}~\shortcite{liu2017deepstyle} attempted to get a better understanding of clothing by utilizing aesthetics and style features, respectively. 
These methods have hammered at comprehending clothing from a holistic perspective via global categories, aesthetics and style information, while failing to dive deep into the details to explore clothing fine-grained semantic aspects. 

However, when purchasing clothing products, it is intuitive that we often have preferences for detailed semantic attributes (such as neckline, heel height, skirt length) in addition to global impressions. For instance, Figure \ref{fig:intro} shows two users' purchase histories. The first user chooses three products in different categories with the same kind of neckline. Therefore, we can infer that she prefers V-neckline clothes. Meanwhile, the second user prefers the pointed toe incorporated with high-heeled shoes as evidenced by the frequent appearance of these two semantic attributes in her chosen products. In fact, clothing is not atomic and a piece of clothing usually consists of multiple attributes, such as neckline, sleeves length. In some cases, people choose a product because it contains the attributes they like. Therefore, the clothing semantic attribute can not only help us generate a comprehensive representation of items, but also make a deep understanding of user preferences. Unfortunately, there are still many unique challenges inherent in designing an effective solution to integrate semantic attribute information into fashion recommendation. On the one hand, it's difficult to obtain clothing semantic attributes features without the manual attribute annotation in the large-scale E-commerce data. On the other hand, the user preferences are sophisticated, while traditional methods usually transform the item image into a latent vector directly. These two aspects make it hard to generate explainable recommendations with current recommendation models.

To address the challenges mentioned above, in this paper, we propose
a novel \textit{S}emantic \textit{A}ttribute \textit{E}xplainable \textit{R}ecommender \textit{S}ystem (SAERS) for fashion recommendation. In SAERS, we introduce a fine-grained interpretable space named semantic attribute space, where each dimension corresponds to a semantic attribute. We project users and items into this space to capture the user's fine-grained preferences and generate explainable recommendations.
Specifically, we firstly develop a Semantic Extraction Network (SEN), which is used to extract the region-specific attribute representations in a weakly-supervised manner. Through SEN, each item is projected to the semantic attribute space. Then, to capture the diversity of the semantic attribute preference among items, we design a Fine-grained Preferences Attention (FPA) module to automatically match the user preference for each attribute in semantic attribute space, and aggregate all attributes with different weights. Finally, we optimize the SAERS in Bayesian Personalized Rank (BPR) framework. Extensive experiments on the large-scale real-world Amazon dataset reveal that SAERS not only significantly outperforms several baselines on the visual recommendation task, but also provides interpretable insights by highlighting attribute semantic in a personalized manner.


\section{Related Work}
Generally, the related work can be grouped into the following three categories, i.e., visual-aware recommendation, attribute localization and explainable recommendation.

\paragraph{Visual-aware Recommendation.} Visual-aware recommendation uses visual signals presented in the underlying data to model visual characteristics of items and user preferences. Some previous works~\cite{mcauley2015image,he2016vbpr,he2016ups,he2016sherlock,HASC2019} directly utilized the ImageNet pre-trained CNN to generate items' visual representation. To further explore the information contained in the product image, Liu \textit{et al.}~\shortcite{liu2017deepstyle} extracted style features to represent item via eliminating the corresponding categorical information from the visual feature vector generated by CNN.  Yu \textit{et al.}~\shortcite{yu2018aesthetic} leveraged an aesthetic network to obtain image aesthetic elements for fashion recommendation task.  In \cite{kang2017visually,lei2016comparative}, Siamese framework was utilized to obtain task-specific visual representation. This framework allowed CNN to be trained comparatively with two `copies' and joined by a triplet loss function. Unfortunately, these methods focused on products' global characteristics and transferred item image into a fixed latent vector. However, the fine-grained attribute has been largely under-exploited.
\begin{figure}
    \centering
    \includegraphics[width=0.45\textwidth, height=100pt]{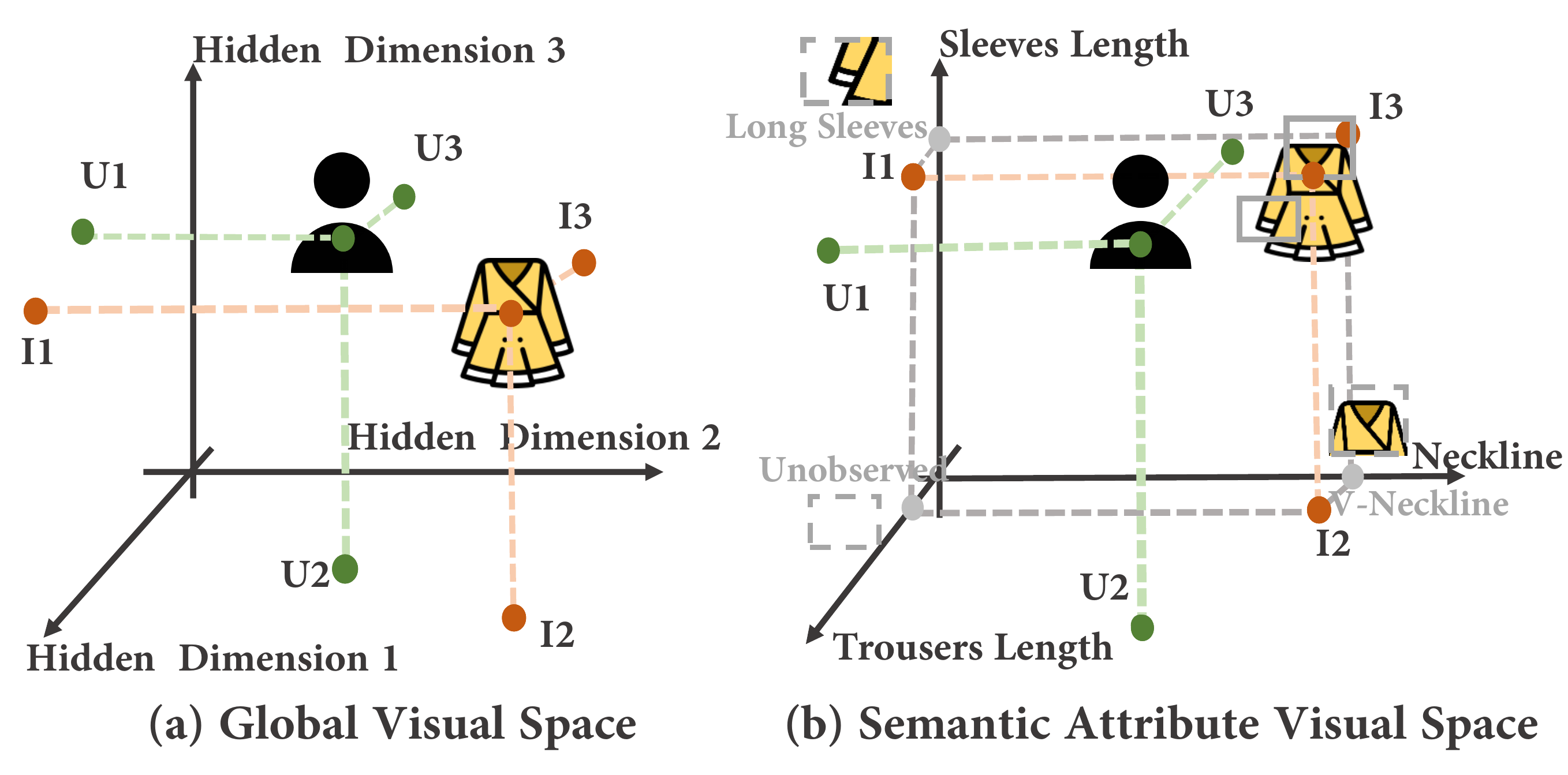}
    \caption{Difference between the conventional (a) Global Visual Space and our (b) Semantic Attribute Visual Space.}
    \centering
    \bigskip
    \label{fig:projection}
\end{figure}
\paragraph{Attribute Localization.} In computer vision, the semantic attribute is a type of important auxiliary information and has been leveraged as a fine-grained representation for image understanding~\cite{10.1007/978-3-642-33712-3_44,kovashka2012whittlesearch}. Researchers have demonstrated that utilizing localized representation to model attributes is more effective \cite{xiao2015discovering}.  Singh and Lee~\shortcite{singh2016end} presented a ranking framework that combines a spatial transformer network to localize attributes. Ak \textit{et al.}~\shortcite{ak2018learning} proposed an efficient weakly-supervised attribute localization and representation method through Class Activation Maps (CAM) \cite{zhou2016learning}, which could highlight the most informative image regions relevant to the predicted class. Nevertheless, in CAM, the fully connected layer of CNN needs to be replaced to global average pooling (GAP) layer and re-trained. For the sake of convenience, in this paper, we use Grid-CAM \cite{selvaraju2017grad} instead to extract attributes' relevant regions.

\paragraph{Explainable Recommendation.} Although conventional recommender systems~\cite{yu2018multiple,zhu2018deep,liu2011personalized,Li:2018:LHP:3219819.3220014} have made great strides, their recommendations are often unexplained and therefore difficult to convince users. To solve this urgent issue, explainable recommendation~\cite{zhang2018explainable,abdollahi2017accurate} has become a very important research direction in recent years. Bilgic \textit{et al.}~\shortcite{bilgic2005explaining} adopted statistical neighbors' ratings and the most impacted bought items to help users understand the recommendations.  McAuley \textit{et al.}~\shortcite{mcauley2013hidden} and Zhang \textit{et al.}~\shortcite{zhang2014explicit} improved the interpretability of collaborative filtering models by utilizing explicit factor based matrix factorization (MF) as explainable components. They aligned each latent dimension in MF with a particular explicit product aspects/topics. In deep learning based methods, researchers discovered that the attention mechanism was capable of improving model explainability by automatically learning the importance of explicit features and also refined user/item embedding \cite{chen2018visually,explainable-recommendation-through-attentive-multi-view-learning,chen2017attentive}.  Chen \textit{et al.}~\shortcite{chen2018visually} proposed an attentive architecture with the supervision of user implicit feedback as well as textual reviews to capture users’ visual preferences. These works attempted to make intuitional explanations for the recommendations but were limited on the item level. In this paper, we take a further step to explain the user preferences on the visual attribute level.

\begin{figure*}
\includegraphics[width=7.0in,height=3.4in]{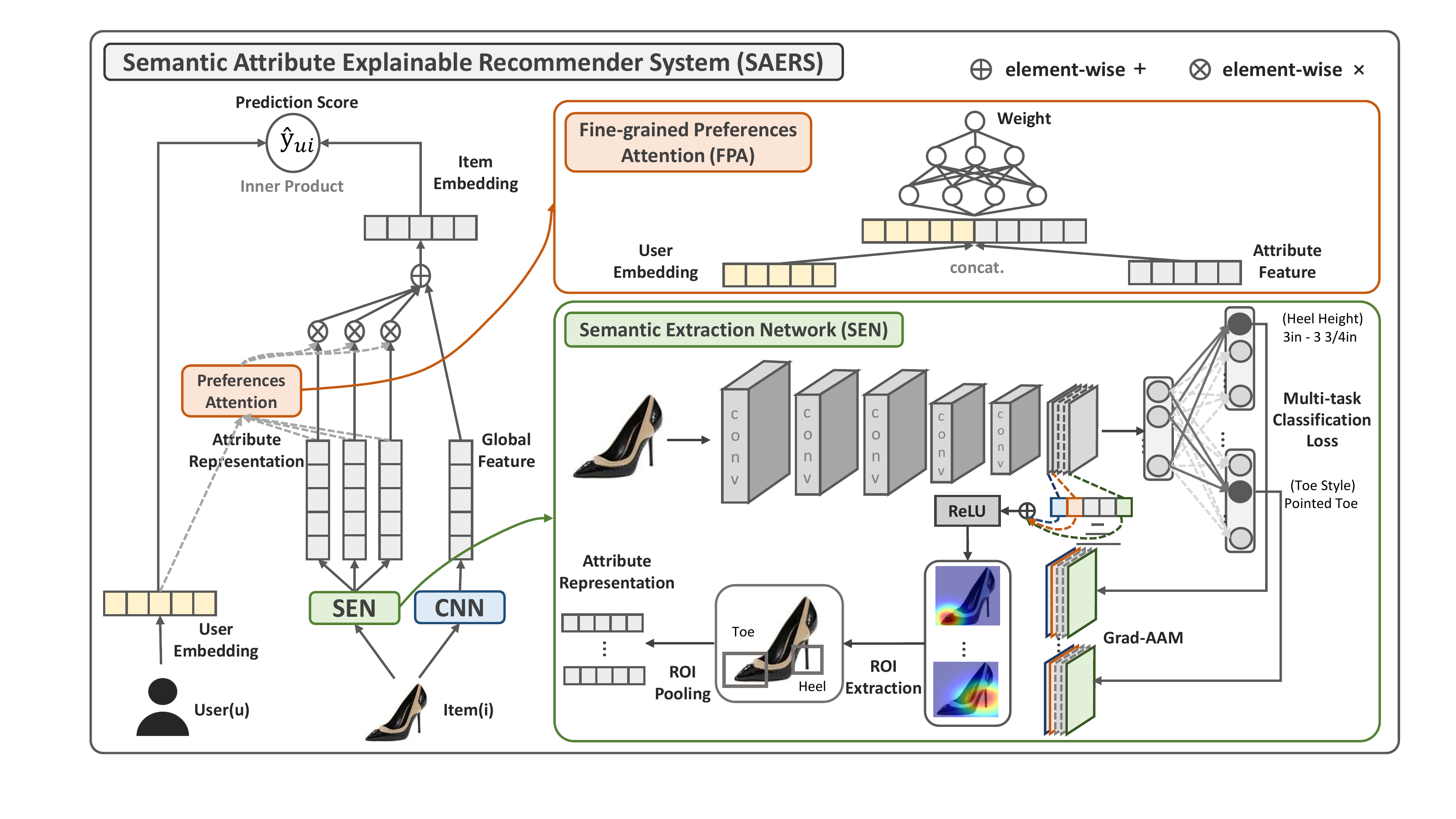}
\caption{The architecture for Semantic Attribute Explainable Recommender System (SAERS)}
\label{fig:framework}
\end{figure*}


\section{SAERS: Semantic Attribute Explainable Recommender System}

In this section, we introduce our proposed SAERS for addressing the personalized fashion recommendation problem. 

Most existing works represent users and items in a global visual space (as shown in Figure \ref{fig:projection}(a)), but the meaning of each dimension is unknown, reducing the interpretability of recommendations.
To bridge this gap, SAERS utilizes a new semantic attribute visual space. In this space, each dimension represents an attribute, which corresponds to different regions of the clothing. As shown in Figure \ref{fig:projection}(b), items are split into several semantic attributes via Semantic Extraction Network, then projected into the semantic attribute visual space. After that, users are projected according to fine-grained preferences for clothing attributes. As shown in Figure~\ref{fig:framework}, SAERS contains two main components, i.e., Semantic Extraction Network (SEN) and Fine-grained Preferences Attention (FPA). Specifically, with SEN, we first obtain the fashion item projections in the semantic feature space. Next, we design FPA to project users into the same semantic feature space. Then, we jointly learn the item representation in both global visual space and semantic attribute visual space under a pair-wise learning framework. Finally, with attribute preference inference, we can generate the explainable recommendations.

\subsection{Projecting Item into Semantic Attribute Space}

In this subsection, we investigate how to project items into semantic attribute space. First, we need to divide the item into several parts (semantic attributes), where each part corresponds to one dimension in the space. And then, we extract attribute representations for them. Whereas, most real-world E-commerce datasets lack attribute annotations to learn semantic attribute representation directly. To this end, we borrow a fine-grained labeled data and pre-train a Semantic Extraction Network (SEN), which is used to extract the region-specific attribute representations and simultaneously locate and classify attributes.

Due to the lack of training data with bounding box annotations, we cannot leverage traditional object detection methods~\cite{girshick2015fast} to locate and extract attribute representations straightforwardly. Therefore, we utilize Gradient-weighted Attribute Activation Maps (Grad-AAM) \cite{selvaraju2017grad} to develop SEN in a weakly-supervised manner. Grad-AAM is capable of producing a coarse location highlighting according to the gradient information of target attribute in CNN. In SEN, we first train a classification CNN. Then, according to the results of CNN, we calculate the Grad-AAM to get attributes location. Finally, the region-specific attribute representations can be obtained by ROI pooling layer.

We combine the \textit{UT-Zap50K shoes\footnote{\url{http://vision.cs.utexas.edu/projects/finegrained/utzap50k/}}} dataset and the \textit{Tianchi Apparel\footnote{\url{https://tianchi.aliyun.com/competition/entrance/231671/information}}} dataset, which contains 50,025 shoe and over 180,000 apparel image-level attribute annotations respectively. As listed in Table \ref{table:attribute describe}, each row represents an attribute. We employ 12 attributes that cover the various locations of the clothing, with each attribute is classified into several classes as shown after the colon. Then, the combined dataset is used to train a multi-task classification network. The network architecture can be arbitrary. In this work, ResNet-50 \cite{he2016deep} is utilized. We use the following classification loss to train the network:
\begin{equation}
     L _ { C } = - \sum _ { I = 1 } ^ { N } \sum _ { a = 1 } ^ { A } \log \left( p \left( \hat{y} _ { I a } | y _ { I a } \right) \right),
\label{equation:classification} 
\end{equation}
where $\hat{y} _ { I a }$ represents the ground truth class of the $a ^ {t h}$ attribute of the $I ^ {t h}$ image. $y _ { I a }$ is the corresponding predicted result. $N$ is the number of training examples and $A$ is the number of attributes. The posterior probability estimates the probability of $y _ { I a }$ to be classified as $\hat{y} _ { I a }$.

The classification loss is used to discover the most relevant regions of attributes. After the classification network has converged, we start to calculate the Grad-AAM $M _ { \text { Grad-AAM } } ^ { a_c }$ for class $c$ of an attribute $a$, where $c$ is determined by the class that maximizes the classification confidence. We first compute the gradient of the score for class $a_c$, $y^{a_c}$~(before the softmax), with respect to feature maps $F ^ { t }$ of the last convolutional layer, \textit{i.e.,} $\frac { \partial y ^ { a_c } } { \partial F ^ { t } }$. These gradients flowing back are global-averaged-pooled to obtain the neuron importance weights $\alpha _ { t } ^ { a_c }$:
\begin{equation}
    \alpha _ { t } ^ { a_c } = \overbrace { \frac { 1 } { Z } \sum _ { m } \sum _ { n } }^{\mathclap{\text{global average pooling}}} \underbrace { \frac { \partial y ^ { a_c } } { \partial F _ { m n } ^ { t } } } _ {\mathclap{\text { gradients via backpop }}}.
\end{equation}

Here, the $(m,n)$ indicates the spatial location of the $t^{th}$ channel feature map $F ^ { t }$ in the last convolutional layer. This weight $\alpha _ { t } ^ { a_c }$ represents a \textit{partial linearization} of the deep network downstream from $F$, and captures the `importance' of feature map $t$ for the target class $a_c$.

We perform a weighted combination of forwarding activation maps, and follow it by a ReLU to obtain,
\begin{equation}
    M _ { \text { Grad-AAM } } ^ { a_c } =  ReLU \underbrace { \left( \sum _ { t } \alpha _ { t } ^ { a_c } F ^ { t }  \right) } _ { \text { linear combination } }.
\label{equation:grad-aam}
\end{equation}

Actually, each element of $M _ { \text { Grad-AAM } } ^ { a_c }$ indicates attribute class ${ a_c }$'s contribution of the activation. Meanwhile, these elements can be associated with the location of the input image. Consequently, by using $M _ { \text { Grad-AAM } } ^ { a_c }$, attribute location can be added to the network. In order to do so, all attribute locations are estimated with a simple hard threshold technique. As the implementation in \cite{ak2018learning}, the pixel values that are above 20\% of the maximum value in the generated map are segmented. This is followed by estimating a bounding box, that covers the largest connected region in the Grad-AAM. This step is repeated for each attribute. With attribute's bounding box and the last convolutional layer's feature map as input, then the ROI pooling layer \cite{girshick2015fast} is used to generate region-specific attribute representation. 

Accordingly, given a clothing image, we can obtain 12 attribute representation through the pre-trained SEN as introduced above. We use $f_a^k$ to denote the $k ^ {t h}$ attribute feature generated by SEN.

\begin{table}[]
\centering
\begin{tabular}{|c|l|}
\hline
\textbf{Category}& \multicolumn{1}{c|}{\textbf{Attribute: Class}}       \\ 
\hline
Top        & \begin{tabular}[c]{@{}l@{}}
                \textit{high neck}: ruffle semi-high, turtle,...        \\                                   
                \textit{collar}: rib collar, puritan collar,...         \\
                \textit{lapel}: notched, shawl, collarless,...          \\ 
                \textit{neckline}: V, square, round,...                 \\ 
                \textit{sleeves length}: sleeveless, cap, short,...     \\ 
                \textit{body length}: high waist, long, regular,...
            \end{tabular}                                               \\ 
\hline
Bottom     & \begin{tabular}[c]{@{}l@{}}
                \textit{skirt length}: short, knee, midi, ankle,...     \\ 
                \textit{trousers length}: short, mid, 3/4, cropped,...
             \end{tabular}                                              \\ 
\hline
Shoes      & \begin{tabular}[c]{@{}l@{}}
                \textit{heel height}: flat, 1 in-7/4 in, under 1 in,... \\ 
                \textit{boots height}: ankle, knee-high, mid-calf,...   \\ 
                \textit{closure}: lace-up, slip-on, zipper,...          \\ 
                \textit{toe style}: round, pointed, peep, open,...
            \end{tabular}                                               \\ 
\hline
\end{tabular}
\caption{List of semantic attributes used in our method}
\label{table:attribute describe}
\bigskip
\centering
\end{table}

\subsection{Projecting User into Semantic Attribute Space}
In this subsection, we describe how to project users into the semantic attribute space and capture fine-grained user preference on attributes.

To get the final embedding of item $i$ with respect to all the attribute features, our first intuition is to equally fuse all the features with an average pooling as follows: 
\begin{equation}
    f ( i ) = \frac { 1 } { A } \sum _ { k = 1 } ^ { A } E^k f_a^k(i),
\label{equation:fuse}
\end{equation}
where $f_a^k (i)\in\mathbb{R}^{m}$ denotes the $k ^ {t h}$ attribute feature of item $i$, $E^k \in\mathbb{R}^{d \times m}$ is a matrix for transferring the $k^{th}$ attribute feature to lower dimension $d$. However, as discussed in the introduction, there may be a variety of clothing attributes that users are interested in. Different attributes may have different impacts on the candidate item $i$ when considering whether user $u$ will buy $i$ or not. Therefore, we argue that the item embedding should be fused with adaptive weights to capture users' dynamic preference over attributes and enhance the interpretability of recommendations. Inspired by the recent success of attention mechanism \cite{chen2018visually}, which allows different parts to contribute differently when compressing them to a single representation. We propose an attribute-level attention module, namely \textit{Fine-grained Preferences Attention} (FPA). Specifically, as illustrated in the right upper part of Figure \ref{fig:framework}, we use a DNN $\mathcal { D }$ (in this work is a two-layer multilayer perceptron) to compute attention weight $\alpha _ { ui } ^ { k }$ with user embedding $f(u) \in \mathbb{R}^d$ and item's $k ^ {t h}$ transferred attribute feature $E^k f_a^k(i)$'s concatenated vector as input:
\begin{equation}
\begin{aligned}
    \alpha _ { ui } ^ { k } &= \operatorname { softmax } \left( \mathcal { D } \left( f(u) , E^k f_a^k(i) \right) \right) \\
    &= \frac { \exp \left( \mathcal { D } \left( f(u) , E^k f_a^k(i) \right) \right) } { \sum _ { k = 1 } ^ { A } \exp \left( \mathcal { D } \left( f(u) , E^k f_a^k(i) \right) \right) }.
\end{aligned}
\label{equation:attention}
\end{equation}

The embedding of item $i$ can thus be calculated as the weighted sum of its attribute features:
\begin{equation}
    f ( i ) = \sum _ { k = 1 } ^ { A } \alpha _ { ui } ^ { k } E^k f_a^k(i).
\end{equation}

Through this way, users are aligned in the semantic attribute space, and the preference towards each attribute can be captured.

Besides the semantic attributes, clothes usually have global characteristics information such as style,  categories and aesthetics. Consequently, 
the embedding of item $i$ can be calculated as the weighted sum of its attribute features plus the global characteristics feature:
\begin{equation}
    f ( i ) = \sum _ { k = 1 } ^ { A } \alpha _ { ui } ^ { k } E^k f_a^k(i) + f_g(i),
\end{equation}
where $f_g(i) \in \mathbb{R}^d$ denotes the global characteristics feature. In this work, we leverage a Siamese CNN architecture based on the AlexNet \cite{krizhevsky2012imagenet} to obtain global characteristics feature.

\subsection{Model Learning Stage}
When making predictions, we feed the user embedding $f(u)\in \mathbb{R}^d$ and the final item embedding $f(i)\in \mathbb{R}^d$ into a function:
\begin{equation}
    \hat { y } _ { u i } = \mathcal { P } \left( f(u) , f(i) \right),
\end{equation}
where $\mathcal{P}$ is an arbitrary prediction function, such as the inner product between the corresponding embeddings, or a prediction neural network \cite{he2017neural}. Here, we choose the inner product $\hat { y } _ { u i } =  f(u) \cdot f(i)$ as a specific implementation, as it gives us better training efficiency and avoids overfitting on our large-scale data.
Since we address the fashion recommendation task from the ranking perspective, we employ a pairwise learning method to optimize model parameters. The assumption of pairwise learning is that the model could predict a higher score for an observed interaction that its unobserved counterparts. We adopt the Bayesian Personalized Ranking (BPR) \cite{rendle2009bpr} loss, which has been widely used in recommender systems:
\begin{equation}
     L  = \sum _ { ( u , i , j ) \in \mathcal { D } _ { S } } - \ln \sigma \left( \hat { y } _ { u i } - \hat { y } _ { u j } \right) + \lambda \| \Theta \| ^ { 2 },
\end{equation}
where $\sigma$ is the sigmoid function and $\lambda$ is the regularization hyper-parameter. $\Theta$ includes all model parameters. $\mathcal { D } _ { S } $ is the training set, which consists of triples in the form $(u, i, j)$. $u$ denotes the user together with an interacted item $i$ and a non-observed item $j$. In each iteration, we sample a user $u$, a positive item $i \in \mathcal { I } _ { u } ^ { + }$, and a negative item $j \in \mathcal { I } \backslash \mathcal { I } _ { u } ^ { + }$. Thus, the convolutional neural network $\Phi$, which is used to extract item's global characteristic feature, has two images: $\mathbf { X } _ { i }$ and $\mathbf { X } _ { j }$ to consider. Both CNNs $\Phi \left( \mathbf { X } _ { i } \right)$ and $\Phi \left( \mathbf { X } _ { j } \right)$ share the same weights. To optimize the above objective function, we employ the widely used Adam \cite{kingma2014adam} optimization algorithm.

\subsection{Attribute Preference Inference} \label{ssec:explanation}
We project users and items into a new interpretable space. In this space, the user's preferences for attributes can be calculated, making it possible to generate personalized explainable recommendations. Specifically, when we recommend an item to a user, the user’s preference towards each attribute of the item can be obtained according to the attention weight (Equation \ref{equation:attention}). The attribute with a large attention weight indicates the user's preference. Utilizing the SEN’s location (Equation \ref{equation:grad-aam}) and classification (Equation \ref{equation:classification}) ability, we can further go back to find out where the attribute is located and what the specific class is. 

Accordingly, the explanation consists of three parts: 
(1) SAERS uses a bounding box to highlight which part of the product image the user might like. 
(2) SAERS provides which semantic attribute the highlighted part belongs to. 
(3) SAERS provides the possibility that the user likes the semantic attribute. We will demonstrate the explainable recommendation results in the experiment section.

\section{Experiments}
In this section, we conduct experiments on a real-world dataset to verify the feasibility of our proposed framework. We first introduce the experimental setup, followed by the experiment results. We will also discuss the effectiveness of semantic attribute features and give a case study of SAERS's visual explainablility.

\subsection{Experimental Setup}

\paragraph{Dataset.} We evaluate our methods on a real-world e-commerce dataset, i.e., Amazon Fashion. This dataset was introduced in~\cite{he2016vbpr,kang2017visually} and consist of reviews of clothing items crawled from \textit{Amazon.com}. Amazon Fashion contains six representative fashion categories (men/women's tops, bottoms and shoes). We treat users' reviews as implicit feedback. There are 45,184 users, 166,270 items, and 358,003 records in total. The sparsity of the dataset is 99.9952\%. In Amazon Fashion dataset, we discard inactive users with purchase records less than 5, similar as previous works \cite{he2016vbpr,kang2017visually}. For each user, we randomly select one record for validation and another one for test. 
\paragraph{Evaluation Protocol.} We evaluate our method in two experiment scenarios: 
(1) In pair-wise recommendation, we use the \textit{Area Under the Roc Curve} (AUC) \cite{rendle2009bpr} to measure the probability that a model will rank a randomly chosen positive instance higher than a randomly chosen negative one.
(2) In list-wise top-\textit{N} recommendation, the \textit{Normalized Discounted Cumulative Gain} (NDCG@\textit{N}) are calculated to evaluate the performance of the baselines and our model. In order to reduce the computational cost, we randomly select a user, and then randomly sample 500 unrated items at each time and combine them with the positive item the user likes in the ranking process. We repeated this procedure 10000 times and report the average ranking results~\cite{wu2017modeling}.
\paragraph{Implementation Details.}  For all the models (except Random and PopRank), we tune hyper-parameters via grid search on the validation set, with a regularizer selected from [0, 0.0001, 0.001, 0.01, 0.1, 1], learning rate selected from [0.0001, 0.001, 0.01], and the latent feature dimension of [10, 30, 50, 100]. We use mini-batch size of 256 to train all the models until they converge. All experiments are trained with NVIDIA K80 graphics card and implemented by \textit{Tensorflow}.

\subsection{Comparison Methods}
To demonstrate the effectiveness of SAERS, we compare it with the following alternative methods:
\begin{itemize}[leftmargin=*,itemsep=2.5pt]
		\setlength{\itemsep}{0pt}
	\setlength{\parsep}{0pt}
	\setlength{\parskip}{0pt}
  \item \textbf{Random:} This baseline ranks items randomly for all users. By definition, this method has AUC=0.5.
  \item \textbf{PopRank:} This is a nontrivial method that items are ranked in order of their popularity within the system.
  \item \textbf{BPR-MF} \cite{rendle2009bpr}\textbf{:} This is a classical matrix factorization method based on  pairwise personalized ranking for implicit feedback datasets.
  \item \textbf{VBPR} \cite{he2016vbpr}\textbf{:} One of the state-of-the-art methods for visually-aware personalized ranking from implicit feedback. VBPR is a form of hybrid content-aware recommendation that makes use of pre-trained CNN features of product images.
  \item \textbf{DeepStyle} \cite{liu2017deepstyle}\textbf{:} A state-of-the-art method for visual recommendation, based on the learned style features and the BPR framework. The style features are obtained by subtracts categories information from visual features of items generated by CNN.
  \item \textbf{JRL} \cite{zhang2017joint}\textbf{:} It is a state-of-the-art neural recommender system, which can leverage multimodal side information for Top-N recommendation. We only use image visual information in this baseline.
\end{itemize}



\begin{figure}
\includegraphics[width=0.47\textwidth,height=105pt]{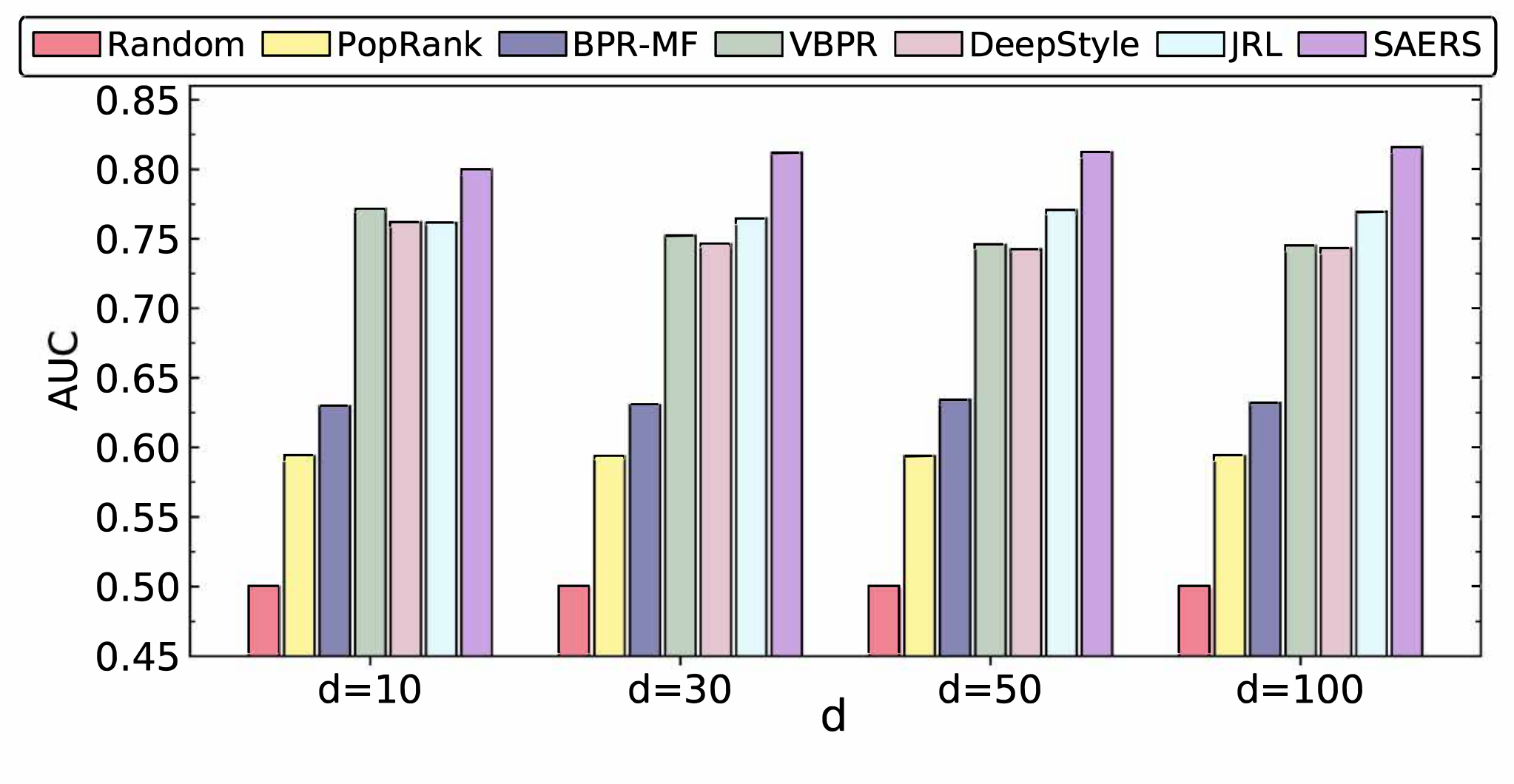}
\caption{Overall performance of difference choices of embedding dimension \textit{d} in \textit{All Items} scenario.}
\label{fig:auc_k_all}
\end{figure}

\begin{figure}
\includegraphics[width=0.47\textwidth,height=105pt]{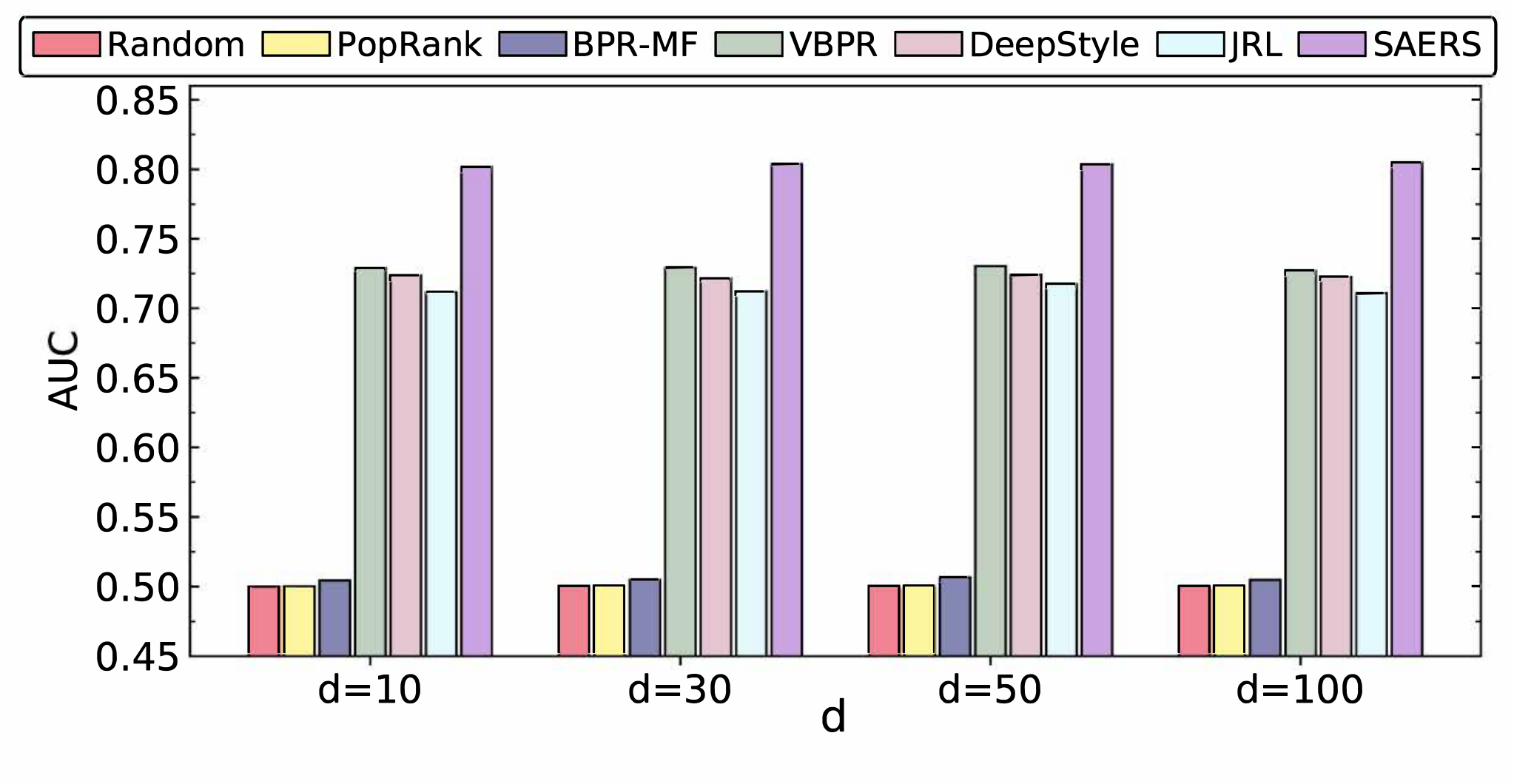}
\caption{Overall performance of difference choices of embedding dimension \textit{d} in \textit{Cold Items} scenario.}
\bigskip
\label{fig:auc_k_cold}
\end{figure}

\subsection{Recommendation Performance}
\begin{figure}
\includegraphics[width=0.48\textwidth]{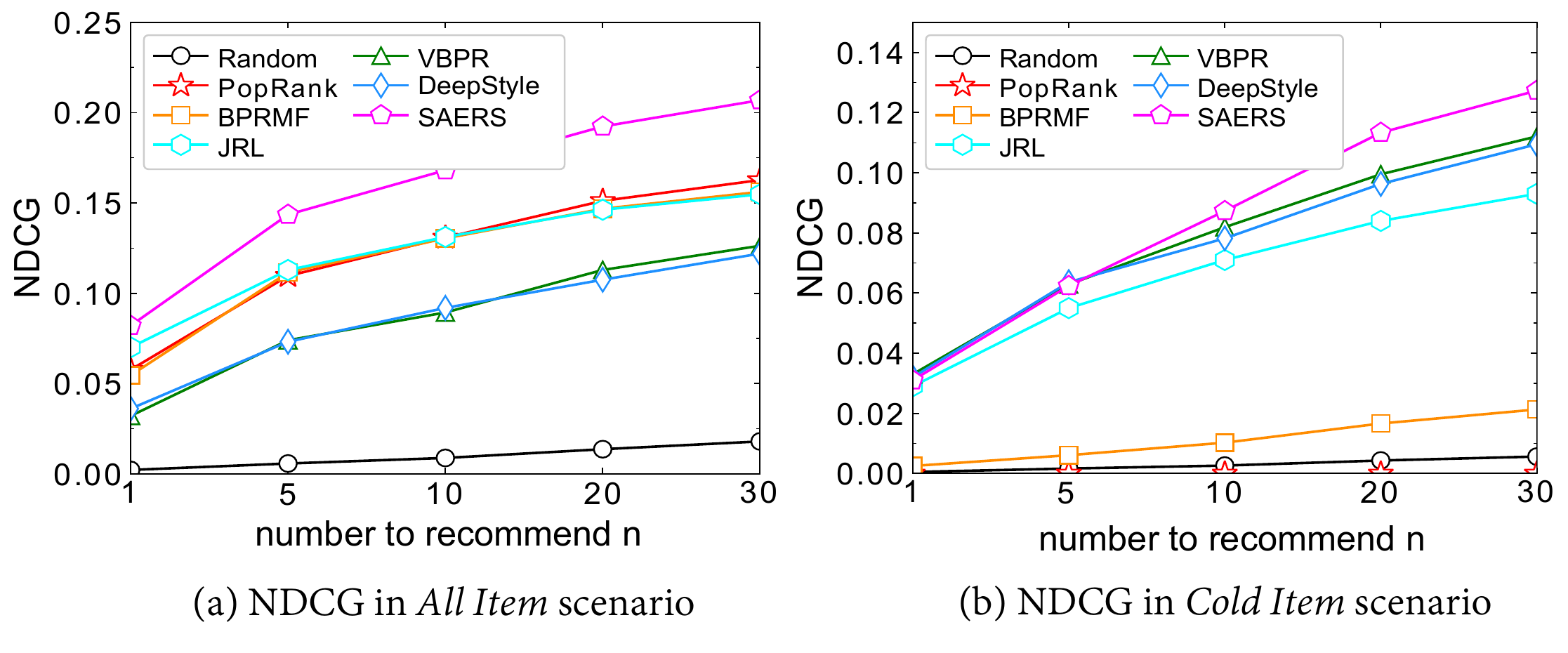}
\caption{NDCG@\textit{N} in top-\textit{N} recommendation for \textit{All Items} and \textit{Cold Items} scenarios.}
\label{fig:ndcg}
\end{figure}

\begin{figure}
\centering
\includegraphics[width=2.8in,height=1.72in]{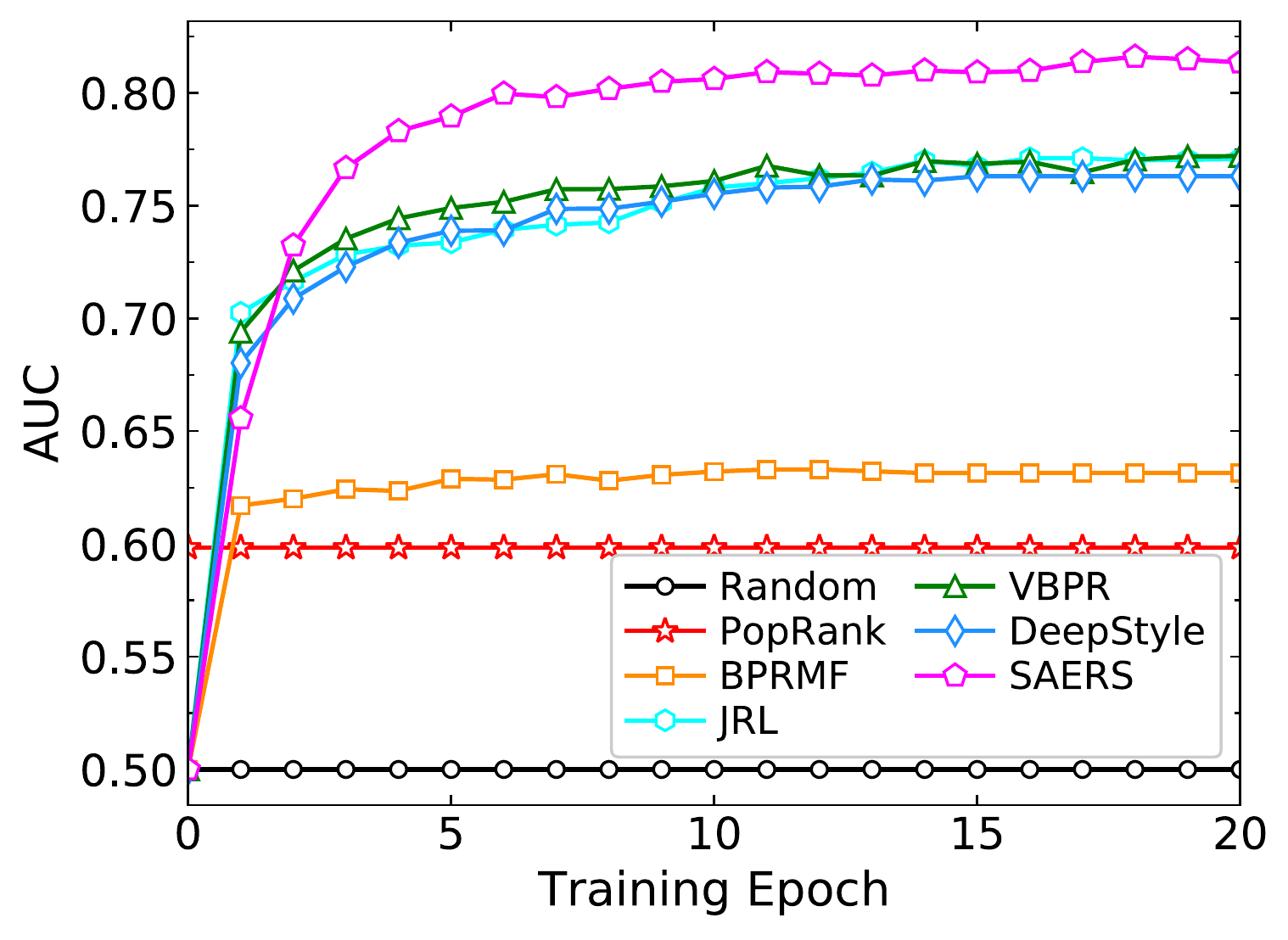}
\caption{Performance of training epochs (test set)}
\label{fig:auc}
\centering
\bigskip
\end{figure}

In this section, we evaluate our model for fashion recommendation. Figure \ref{fig:auc_k_all} and Figure \ref{fig:auc_k_cold} present the recommendation performance on AUC in \textit{All Items} and \textit{Cold Items} scenarios, respectively. For the latter, we seek to estimate relative preference scores among items that have never been observed at training time. Figure \ref{fig:ndcg} depicts the Top-\textit{N} recommendation performance on two scenarios when embedding dimension \textit{d}=100. Figure \ref{fig:auc} shows the AUC during training. The main findings are summarized as follows:
\begin{itemize}[leftmargin=*,itemsep=2.5pt]
	\setlength{\itemsep}{0pt}
	\setlength{\parsep}{0pt}
	\setlength{\parskip}{0pt}
\item We can find that our proposed SAERS achieves the best performance on AUC, with the improvement by up to 5.8\% compared to the second-best method across all datasets, and 10.2\% in cold-start scenarios. In Figure \ref{fig:ndcg}, we can also find that our model outperforms the other baseline methods in Top-\textit{N} recommendation at two scenarios. The result demonstrates the significant benefits of incorporating semantic attribute features in fashion recommendation.
\item Our method surpasses the strongest collaborative filtering based method (BPR-MF) significantly, especially in \textit{Cold Items} scenarios. This phenomenon is largely due to the fact that the dataset is particularly sparse. The sparsity of the dataset is 99.9952\% and on average each item is only be purchased 2.15 times. This further illustrates the importance of using content-based recommendation methods and in-depth understanding of image information in fashion recommendation tasks.
\item The performance of DeepStyle using style information does not work better than VBPR, by directly using CNN pre-trained features. DeepStyle assumes that style feature can be obtained by subtracting categorical information from visual features of items generated by CNN. However, in fashion recommendation tasks, there are few categories of clothing, but the style is indeed diverse. Therefore, the assumption could not work well. This reflects the difficulty of effectively incorporating content information into the recommender system.
\item Figure \ref{fig:auc_k_all} and \ref{fig:auc_k_cold} show the sensitivity of the models to the embedding dimension \textit{d}. It shows that SAERS is robust to hyper-parameter tuning.
\end{itemize}

\subsection{Necessity of the Semantic Attribute}
In this subsection, we discuss the necessity of using the semantic attribute features extracted by SEN. We construct three variant implements based on VBPR~\cite{he2016vbpr} and SAERS to verify the effectiveness of incorporating the semantic attribute information.
\begin{itemize}[leftmargin=*,itemsep=2.5pt]
		\setlength{\itemsep}{0pt}
	\setlength{\parsep}{0pt}
	\setlength{\parskip}{0pt}

    \item \textbf{SAFo:} This is a model with \textbf{S}emantic \textbf{A}ttribute \textbf{F}eatures \textbf{o}nly. Items are represented as the fusion of all semantic attributes representations, as shown in Equation \ref{equation:fuse}.
    \item \textbf{VBPR+SAF:} Add the \textbf{S}emantic \textbf{A}ttribute \textbf{F}eatures to the CNN features VBPR used as the item representation.
    \item \textbf{SAERS-SAF:} A variant of our model that uses only global characteristic features extract by Siamese CNN to represent items.
\end{itemize}

As shown in Table \ref{table:attribute}, the usage of SAF can increase the effect of VBPR and our model by 1.4\% and 1.8\%, respectively. Experimental results show our SAERS model, which captures both local attribute semantic information and global characteristic information, performs the best result on the \textit{Amazon} dataset since these two kinds of information mutually enhance each other to a certain extent. 
\begin{table}[]
\begin{center}
\begin{tabular}{lcc}
\hline
Models     & AUC             & \multicolumn{1}{l}{Improvement} \\ \hline
SAFo       & 0.7569          & -                            \\
\hline
VBPR       & 0.7718          & -                               \\
VBPR + SAF & 0.7826          & 1.4\%                          \\ \hline
SAERS - SAF & 0.8017          & -                               \\
SAERS       & \textbf{0.8161} & \textbf{1.8\%}                  \\ \hline
\end{tabular}
\end{center}
\caption{Improvement of using semantic attribute feature.}
\bigskip
\label{table:attribute}
\end{table}

\subsection{User Attribute Interest Visualization}
\begin{figure}
\includegraphics[width=0.48\textwidth]{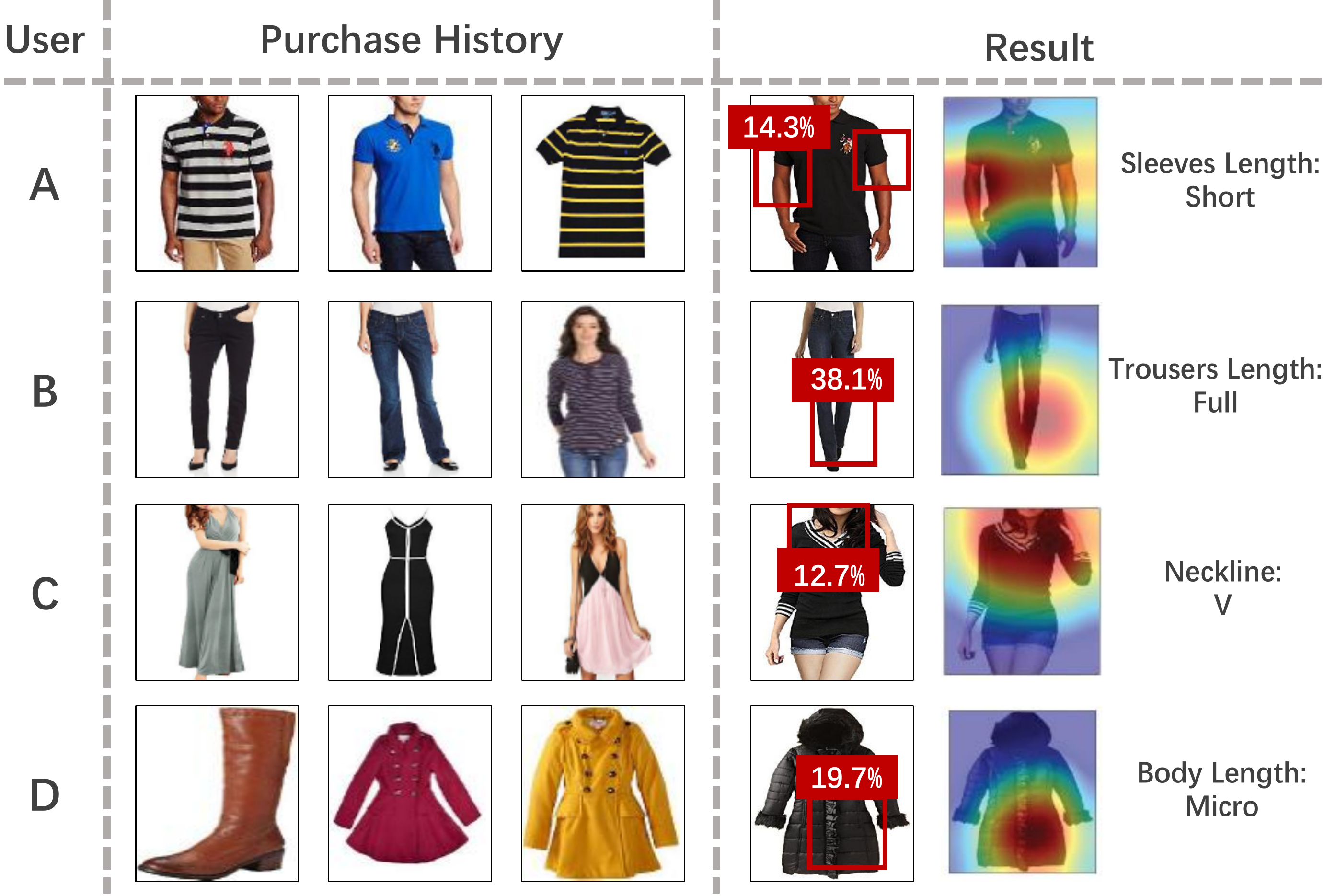}
\caption{Examples of the visual explanations (The IDs of User A, B, C and D are A2ZEFKXUBN2IIS, A2OLJU8M0L4YZD, AIUZT213W0WDE and AQ6BC76U77C7X)}
\bigskip
\label{fig:case_study} 
\end{figure}

In order to better demonstrate the interpretability of our model, we randomly selected four users. As shown in Figure \ref{fig:case_study}, for each user, we present 3 logs from the purchase history as well as our recommendation result. The value in the red box indicates the attention weight. The greater weight demonstrates the higher personal preferences for this attribute. We show the user's favorite attribute calculated by the Fine-grained Preferences Attention (FPA) and locate it via Grad-AAM.
(1) Attribute class, (2) attribute location and (3) attention weight are corresponding to three visual explanations we mentioned in Section \ref{ssec:explanation}.
For example, SAERS recommends a T-shirt to user A and tells him that the reason for the recommendation is short sleeves and the position corresponding to the sleeve is highlighted. This is reasonable because the user has purchased three short sleeves clothing. For user C, SAERS recommends a shirt with V-neckline because it accurately captures the user's preference on this attribute. In the recommendation results, we can see that the neckline of this shirt has the largest attention weight, which accounts for 12.76\%. This explains the reason we recommend this shirt: user C likes this kind of neckline.

\section{Conclusions}
In this paper, we proposed SAERS, aiming at comprehending user preference in fine-grained attribute level and providing visual interpretability in fashion recommendation tasks. To achieve this goal, we first introduced a fine-grained interpretable space, namely, semantic attribute space. Then we developed a Semantic Extraction Network (SEN) and Fine-grained Preferences Attention (FPA) module to project users and items into this space, respectively. Finally, we jointly learned the item representation in both global visual space and semantic attribute space under a pair-wise learning framework. The experimental results on a large-scale real-world dataset clearly demonstrated the effectiveness and explanatory power of SAERS.

\section*{Acknowledgements}
This research was partially supported by grants from the National Key Research and Development Program of China (No. 2016YFB1000904) and the National Natural Science Foundation of China (Grants No. U1605251, 61727809, 61602147). 

\bibliographystyle{named}
\bibliography{ijcai19}

\end{document}